\documentclass[letterpaper]{article}
\usepackage{spconf,amsmath,graphicx}
\usepackage{adjustbox}
\usepackage{multirow}
\usepackage{amsmath}
\usepackage{amssymb}
\usepackage[table]{xcolor}
\usepackage{mathrsfs}
\usepackage{dsfont}
\usepackage{bm}
\usepackage{float}
\usepackage{multirow}
\usepackage{booktabs}
\usepackage{nicefrac}
\usepackage{nth}
\usepackage[bf,small,skip=0pt]{caption}
\usepackage{subcaption}
\usepackage{wrapfig}
\usepackage{dsfont}
\usepackage{tikz}
\usetikzlibrary{shapes,arrows,backgrounds,positioning}
\usepackage{mathtools}
\usepackage{dsfont}
\usepackage{adjustbox}
\usepackage{multirow}
\usepackage{url}
\usepackage{color}
\usepackage{wrapfig}
\usepackage{mathrsfs}
\usepackage{ctable}
\usepackage{amsthm}

\usepackage[draft]{todonotes}

\DeclareMathOperator*{\argmax}{arg\,max}
\DeclareMathOperator*{\argmin}{arg\,min}

\newcommand{\xu}[1]{{\color{cyan}{}}}
\newcommand{\ding}[1]{{\color{black}{#1}}}

\newcommand{\mn}[1]{{\color{black}{#1}}}

\title{VoteNet++: Registration Refinement for Multi-Atlas Segmentation}

\name{Zhipeng Ding, Marc Niethammer}
\address{Department of Computer Science, UNC Chapel Hill, USA}
\begin{document}
%
\maketitle
\begin{abstract}
\mn{Multi-atlas segmentation (MAS) is a popular image segmentation technique for medical images. In this work, we improve the performance of MAS by correcting registration errors before label fusion. Specifically, we use a volumetric displacement field to refine registrations based on image anatomical appearance and predicted labels. 
  \ding{We show the influence of the initial spatial alignment as well as the beneficial effect of using label information for MAS performance.} Experiments demonstrate that the proposed refinement approach improves \ding{MAS performance} on a 3D magnetic resonance dataset of the knee.}
\end{abstract}
%

\section{Introduction}
\label{sec:introduction}
\ding{Even though deep segmentation networks~\cite{cciccek20163d} have been highly successful and hence have become very popular, multi atlas segmentation (MAS) benefits from its ability to maintain consistent anatomical structures~\cite{ding2019votenet}.}
Assuming label segmentation strongly associates with image appearance, \mn{atlas-based image segmentation uses one or multiple image registrations to transfer atlas labels from one or multiple images to a target image that is to be segmented. 
MAS obtains a consensus optimization via a label fusion step and has achieved great success in medical image segmentation due to its reliable performance~\cite{iglesias2015multi}. In particular, as multiple atlases are used, individual registration failures are not critical. Nevertheless, reliable image registrations for label propagation are desirable to accurately align anatomical structures across target images. In consequence, better registration can lead to improved MAS performance~\cite{iglesias2015multi}.}

However, registration will in general be inaccurate due to anatomical complexities and variations across different images. As pointed out in~\cite{wang2012multi}, registration errors (i.e., failure by the registration algorithm to correctly recover correspondences between images) are the principal source of error in multi-atlas segmentation approaches. Different strategies to compensate for these errors have been proposed. Patch-based methods~\cite{coupe2011patch,wang2012multi,bai2013probabilistic,xie2019improving} search image neighborhoods at registered atlas positions to find the patches best matching the target image. Segmentation performance can then be moderately improved by using the displaced patch when computing the consensus segmentation of the target image~\cite{wang2012multi}. Such a local patch search technique can be viewed as refining the voxel-to-voxel correspondences computed by registration, while relaxing the regularization constraints that registration imposes on deformation fields~\cite{wang2012multi}. The key drawback of this approach is that refinement without spatial regularization tends to produce unrealistic atlas images, thus breaking the consistent appearance of anatomical structures. Since atlas appearance is not directly used for the final label fusion, the consequences of unrealistic atlas registrations are not well studied.
Fig.~\ref{fig:result} shows an example of such an unrealistic atlas appearance after patch-based refinement.

Besides approaches to improve spatial correspondences, different label fusion strategies have been proposed for MAS, \ding{the simplest being plurality voting~\cite{heckemann2006automatic} and global or local weighted voting~\cite{artaechevarria2008efficient,wang2012multi}}.
Deep learning based approaches for label fusion have recently been proposed. For example, the VoteNet~\cite{ding2019votenet,ding2020votenet+} approaches use a deep convolutional neural network to locally select the most trustworthy atlases and achieve good performance even when followed by label fusion via simple plurality voting. The VoteNet approaches do not directly cope with registration errors, but instead aim at removing atlases with big registration errors from the label fusion step.

In this work, we integrate a registration refinement step into VoteNet+~\cite{ding2020votenet+}. Different from patch based methods~\cite{coupe2011patch,wang2012multi,bai2013probabilistic,xie2019improving}, we estimate a spatially smooth voxel-to-voxel correspondence to retain realistic anatomical atlas appearance after refinement. Specifically, we make use of regularized spatial transformation networks~\cite{jaderberg2015spatial} to obtain a volumetric displacement field to refine atlas-to-target registrations. The refined results are input into VoteNet+~\cite{ding2020votenet+} to obtain a final consensus segmentation. In contrast to prior work on registration refinement for MAS~\cite{bai2013probabilistic}, \ding{the key differences are: 1) we incorporate consistent predicted segmentations while \cite{bai2013probabilistic} uses MAS fused segmentations iteratively to refine registrations;
  2) we use a volumetric displacement field while~\cite{bai2013probabilistic} uses local patch search. Our approach can keep anatomical structures while local patch search would create unrealistic atlas appearance that can not be used in the VoteNet+ framework. 3) we base our approach on deep learning~\cite{ding2020votenet+} while \cite{bai2013probabilistic} uses a Bayesian approach for label fusion. Since~\cite{ding2019votenet} already demonstrated that VoteNet can outperform a patch based approach~\cite{bai2013probabilistic} and local patch search would create unrealistic atlas appearance that can not be used in the VoteNet+~\cite{ding2020votenet+}, we omit direct comparisons with these approaches and instead focus on refining registrations under VoteNet+ framework.}
\noindent
\ding{Our main \textbf{contributions} are as follows: 1)~\emph{Novel Registration Refinement in MAS}: We propose using anatomical appearance \emph{and} predicted labels to refine atlas-to-target registrations via a volumetric displacement field and produce smooth and accurate registration refinements. (2)~\emph{Comprehensive analysis}: We thoroughly examine the effect of registration refinement for MAS with different registration accuracy.}


\section{Methodology}
\label{sec:method}
\begin{figure}[t]
\centering
  \includegraphics[width=\linewidth]{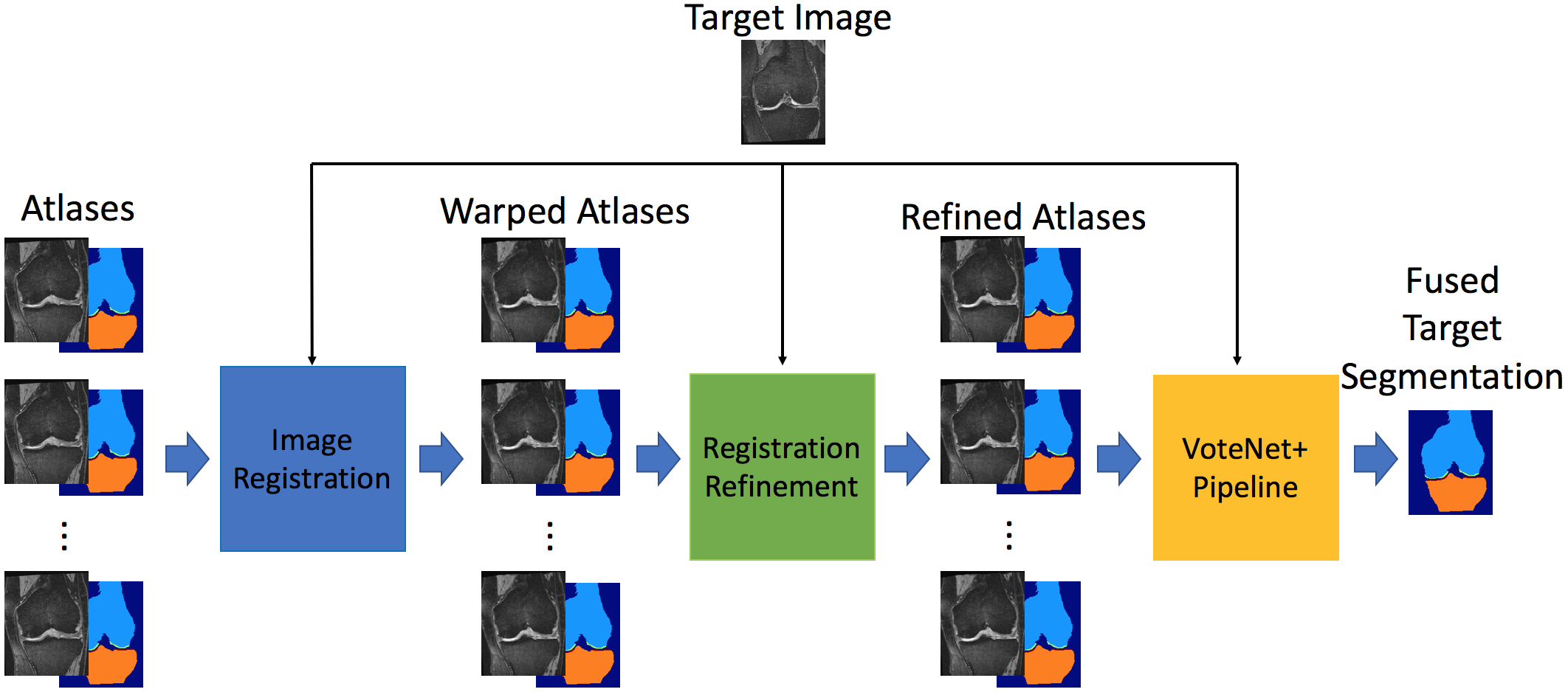}
  \caption{Framework of our VoteNet+ based MAS with refinement step. First, images are registered to target images via an image registration method. Second, a refinement step improves registrations. Lastly, VoteNet+ uses the refined atlases and the target image to obtain a fused consensus target segmentation.}
  \label{fig:framework}
  \vspace{-0.5ex}
\end{figure}

\mn{Fig.~\ref{fig:framework} illustrates our framework. For completeness, Sec.~\ref{sec:mas_overview} briefly introduces the VoteNet+ approach. Sec.~\ref{sec:refine_regs} describes our proposed refinement approach. Necessary preregistration are performed via deep registration networks for speed (see Sec.~\ref{sec:exp_setting} for details).}

\subsection{MAS Overview}
\label{sec:mas_overview}
Let $T_I$ represent the target image that needs to be segmented. Denote the pairs of $n$ atlas images, $A^i_I$, and their corresponding manual segmentations, $A^i_S$, as $A^1 = (A^i_I, A^i_S), A^2 = (A^2_I, A^2_S), ..., A^n = (A^n_I, A^n_S)$. MAS first employs a reliable deformable image registration method to warp all atlas images into the space of the target image $T_I$, obtaining the registered atlas images and their segmentations $\tilde{A}^i = (\tilde{A}^i_I, \tilde{A}^i_S), i=1,...,n$. Each $\tilde{A^i_S}$ is considered as a candidate segmentation for $T_I$. A label fusion method $\mathscr{F}$ produces the final estimated segmentation $\hat{T}_S$ for $T_I$:
\begin{equation}
\label{eq:est}
\hat{T}_S = \mathscr{F}(\tilde{A}^1, \tilde{A}^2, ..., \tilde{A}^n, T_I).
\end{equation}

Furthermore, each atlas may exhibit different anatomical variations. Thus, it is sensible to assign different weights to different atlases~\cite{iglesias2015multi,artaechevarria2008efficient,wang2012multi} to focus on atlases which are the best fits for the target images. Weighting can be global or local. \ding{VoteNet+~\cite{ding2020votenet+} uses deep learning to provide voxel-wise weight assignments and label fusion.} Given a warped atlas image $\tilde{A}_I$ and a target image $T_I$ as inputs, VoteNet+ predicts the probability of $p(\tilde{A}_S = T_S | \tilde{A}_I, T_I)$ at each voxel location. These probabilities are then used for \emph{joint label fusion (JLF)}~\cite{wang2012multi}. JLF accounts for the possibility that atlases may make correlated errors and \ding{computes the weights of atlases via error minimization.}
\ding{Details can be found in \cite{ding2020votenet+}.}
Finally, the segmentation for target image $T_I$ is obtained by solving 
\begin{equation}
\label{eq:goal}
\hat{T}_S(x) = \argmax_{l \in \Omega} \sum_{i=1}^{n} w^i_x \cdot \mathds{1}[\tilde{A}^i_S(x) = l]\,,
\end{equation}
where $l \in \Omega=\{0,\dots,K\}$ is the set of labels ($K$ structures; 0 indicating background), $\mathds{1}[\cdot]$ is the indicator function, and $w^i_x$ is the \ding{$i$-th atlas weight obtained from JLF.}


\begin{figure*}[!t]
\centering
 \includegraphics[width=0.9\linewidth]{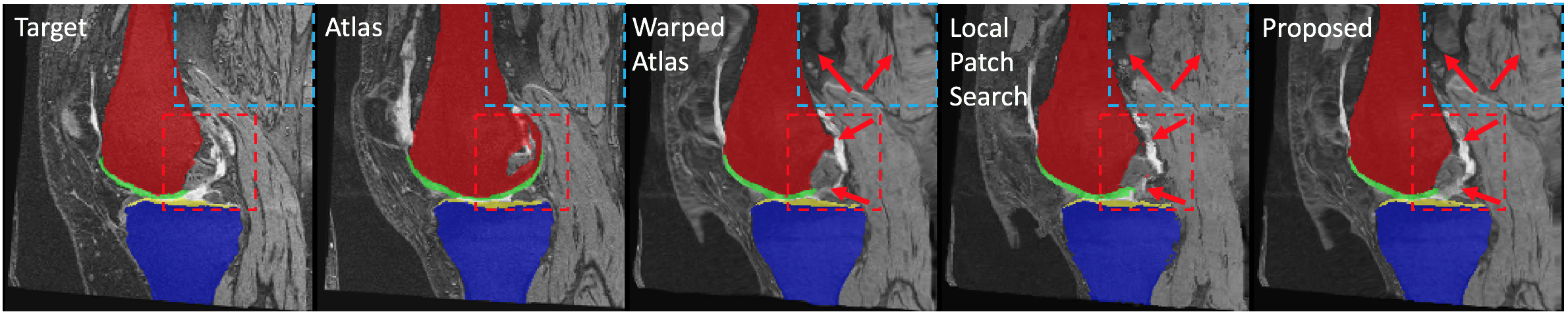}
 \caption{Registration refinement performance. Color indicates different anatomical structures. Local Patch Search results in unrealistic images (blue dotted rectangle) and labels are not smooth at the structure boundaries (red dotted rectangle). Our proposed method maintains realistic anatomical structures and smooth boundaries.}
 \label{fig:result}
 \vspace{-0.5ex}
\end{figure*}

\subsection{Refining Registrations by Numerical Optimization}
\label{sec:refine_regs}
Having introduced the basic background of VoteNet+ based MAS, this section discusses how we refine the atlas-to-target registrations via a volumetric displacement field to obtain more accurate MAS consensus segmentation results.

\noindent
\textbf{Objective Function.} Our goal it to optimize a displacement field to refine the alignment of an already pre-aligned atlas with respect to the target image that we want to segment. We hypothesize that better label fusion by VoteNet+ can be achieved by refining this alignment. We use three different losses for this refinement: an image \ding{dis-}similarity loss, $\mathcal{L}_{img}$;  a segmentation \ding{dis-}similarity loss, $\mathcal{L}_{seg}$; and a loss encouraging spatial regularity $\mathcal{L}_{reg}$. This is essentially a standard registration setup with a similarity measure taking into account images and their segmentations. Note that we do not know the segmentation of the target image, as this is precicely what we are after; we will therefore use a simpler automatic segmentation method to obtain segmentation surrogates (see details below). Specifically, our \ding{optimized displacement field is}
\begin{multline}
\label{eq:loss_function}
f_i^* = \argmin_{f_i} \mathcal{L}_{img} (\tilde{A}^i_I \circ \phi_i^{-1}, T_I) + \alpha \mathcal{L}_{seg} (S_{src}^{i} \circ \phi_i^{-1}, S_{tar}) \\
+ \gamma \mathcal{L}_{reg} (f_i),\quad\text{s.t.}~\phi_i^{-1}(x) = x + f_i(x),
\end{multline}
where $S_{src}$ and $S_{tar}$ are predicted segmentations of $\tilde{A}^i_S$ and $T_I$ respectively \xu{(Why do you use the predicted segmentation for $\tilde{A_S^i}$ here? When you register the atlas to target for the first time, you should have the warped segmentation as well?)\ding{This is for helping to refine registrations. We don't know the segmentation of target image}}; \ding{$\alpha = 3, \gamma = 20,000$ in our experiment} \xu{(Maybe you want to mention $\alpha$ and $\gamma$ later in the experiment section. Up to here, I don't know what $L_{reg}$ is)\ding{Previously, it was specified in experiments, but since I am running out of space, I directly put it here to save space.}} and $\phi_i^{-1}$ is the spatial transformation map. 
$\mathcal{L}_{img}$ encourages a warped atlas image to look similar to the target image; $\mathcal{L}_{seg}$ incorporates a consistent segmentation prediction result into the registration in order to improve registration accuracy; and $\mathcal{L}_{reg}$ encourages smoothness of the transformation. \xu{(You have talked out these three losses above. No need to mention them again.)\ding{This is more detail than the above-mentioned.}}  




We use normalized cross correlation (NCC) as the image similarity measure
\begin{equation}
\label{eq:img_loss}
\mathcal{L}_{img} (\tilde{A}^i_I \circ \phi_i^{-1}, T_I) =  1- NCC(\tilde{A}^i_I \circ \phi_i^{-1}, T_I)\,. 
\end{equation} 

Different from~\cite{bai2013probabilistic} where \ding{the output consensus segmentation of MAS} is used within the registration to improve registration accuracy, we use predicted segmentations (obtained via a deep network, e.g., a U-Net~\cite{cciccek20163d}) for $\mathcal{L}_{seg}$. The idea is that even if segmentations are imperfect they can be reliably used for registration as long as they are reasonably consistent. Specifically, we use a soft multi-class Dice loss
\begin{equation}
\label{eq:seg_loss}
\mathcal{L}_{seg} (S_{ref}, S_{tar}) =  1- \frac{1}{K} \sum_{k=1}^K \frac{\sum_x S_{ref}^{(k)}(x)S_{tar}^{(k)}(x)}{\sum_x S_{ref}^{(k)}(x) + \sum_x S_{tar}^{(k)}(x)}, 
\end{equation} 
where  $k$ indicates a segmentation label (out of $K$), $x$ is the voxel position \xu{(Is the $p$ here the same as $x$ in equation ~\eqref{eq:loss_function}?)\ding{yes}}, and $S_{ref} = S^{i}_{src} \circ \phi_i^{-1}$ is the refined predicted warped atlas segmentation.

\noindent
We use the same smoothness term as in \cite{xu2019deepatlas}, which amounts to regularization based on a bending energy~\cite{rueckert1999nonrigid}. I.e., 
\begin{equation}
\label{eq:flow_loss}
\mathcal{L}_{reg} (f_i) = \frac{1}{M} \sum_{x} \sum_{i=1}^d || H(f_i(x))||^2_{F}, 
\end{equation} 
where $||\cdot||_F$ denotes the Frobenius norm, $H(f_i(x))$ is the Hessian of the $i$-th component of $f$ at position $x$, and $d$ denotes the volumetric dimension ($d = 3$ in our case). $M$ denotes the number of voxels.

\begin{figure}[!b]
\centering
  \includegraphics[width=0.9\linewidth]{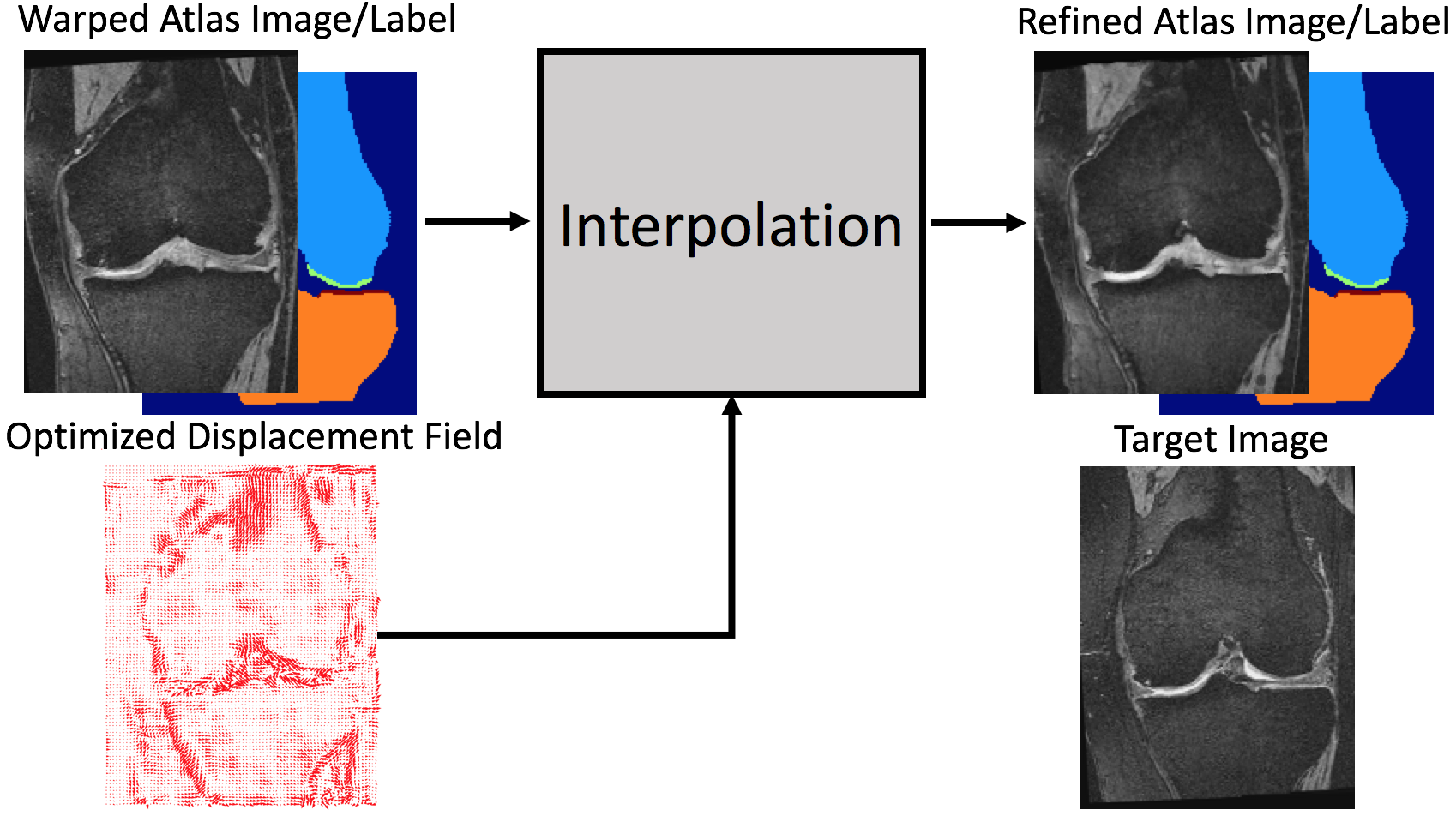}
  \caption{Volumetric transformation. Warped atlas image (label) is transformed via a displacement field using interpolation.}
  \label{fig:stn}
  \vspace{-0.5ex}
\end{figure}

\noindent
\ding{\textbf{Volumetric displacement field.} Fig.~\ref{fig:stn} illustrates the refinement process. Using the optimized displacement field from Eq.~\ref{eq:loss_function} via differentiable trilinear interpolation~\cite{jaderberg2015spatial}, we can refine the warped atlas image to be closer to target image. Thus, the warped atlas segmentation is also expected to be closer to the true target segmentation. Note that we use the warped atlas segmentation here, not the predicted segmentations used in Eq.~\ref{eq:seg_loss} to optimize the displacement field.}
\xu{(Why do you specifically mention "volumetric"? You just need to mention 3D somewhere and it will save you some spaces.)\ding{depends on the final paper length, might delete the volumetric. But keep it is for consistency with spatial transformation network.}}

\begin{table*}[!t] 
\centering
\begin{adjustbox}{max width=0.8\textwidth}
\begin{tabular}{clcccccccc}
\specialrule{.15em}{.05em}{.05em}  
\multicolumn{1}{c}{\multirow{2}{*}{\shortstack{Atlas-to-Target \\ Registration}}} & \multicolumn{1}{l}{\multirow{2}{*}{Method}} & \multicolumn{2}{c}{ASD (mm) $\downarrow$} & \multicolumn{2}{c}{SD (\%) $\uparrow$} & \multicolumn{2}{c}{95MD (mm) $\downarrow$} & \multicolumn{2}{c}{VD (\%) $\uparrow$}\\
\cmidrule(lr){3-4}
\cmidrule(lr){5-6}
\cmidrule(lr){7-8}
\cmidrule(lr){9-10}
\multicolumn{1}{c}{}&\multicolumn{1}{r}{}& \textit{Bone} & \textit{Cartilages} & \textit{Bone} & \textit{Cartilages} & \textit{Bone} & \textit{Cartilages} & \textit{Bone} & \textit{Cartilages} \\
\specialrule{.15em}{.05em}{.05em}
-- &U-Net (baseline)&0.24(0.13)&0.27(0.06)&96.00(1.87)&94.20(2.61)&0.93(0.95)&1.00(0.31)&98.07(0.34)&81.66(3.12)\\
\hline
\hline
\multicolumn{1}{c}{\multirow{4}{*}{Affine}}&PV&1.40(0.24)&1.22(0.32)&44.32(8.10)&55.61(11.80)&3.83(0.66)&3.66(0.75)&90.53(2.12)&33.93(10.95)\\
&VoteNet$_0$+&0.38(0.12)&0.50(0.18)&88.73(6.11)&84.74(9.22)&1.68(0.73)&1.88(0.68)&97.09(0.99)&71.56(9.10)\\
&VoteNet$_0$+ (I)&0.42(0.23)&0.36(0.09)&91.98(3.55)&90.76(4.05)&2.08(1.65)&1.45(0.49)&97.49(0.71)&77.55(4.51)\\
&VoteNet$_0$+ (IS)&\textbf{0.28(0.06)}&\textbf{0.32(0.10)}&\textbf{93.38(2.92)}&\textbf{91.91(4.64)}&\textbf{1.20(0.37)}&\textbf{1.27(0.50)}&\textbf{97.79(0.44)}&\textbf{78.69(5.03)}\\
\hline
\multicolumn{1}{c}{\multirow{4}{*}{DI}}&PV&0.57(0.11)&0.47(0.12)&79.63(5.45)&86.46(5.20)&2.26(0.47)&1.95(0.60)&95.75(1.00)&73.20(5.20)\\
&VoteNet$_1$+&0.25(0.04)&0.32(0.08)&94.58(2.68)&92.20(3.64)&0.93(0.19)&1.23(0.42)&97.86(0.42)&79.13(4.01)\\
&VoteNet$_1$+ (I)&\textbf{0.24(0.04)}&0.32(0.08)&94.72(2.65)&92.35(3.60)&0.92(0.20)&1.21(0.43)&97.88(0.41)&79.29(4.00)\\
&VoteNet$_1$+ (IS)&0.26(0.13)&\textbf{0.28(0.06)}&\textbf{95.62(2.36)}&\textbf{93.94(2.85)}&\textbf{0.90(0.13)}&\textbf{1.03(0.34)}&\textbf{98.01(0.34)}&\textbf{81.04(3.38)}\\
\hline
\multicolumn{1}{c}{\multirow{4}{*}{DIS}}&PV&0.31(0.05)&0.28(0.06)&91.22(3.19)&93.91(2.91)&1.18(0.22)&1.03(0.34)&97.42(0.46)&81.89(3.34)\\
&VoteNet$_2$+&0.22(0.03)&0.26(0.06)&95.55(2.16)&94.56(2.71)&0.85(0.15)&0.96(0.31)&98.04(0.32)&82.26(3.29)\\
&VoteNet$_2$+ (I)&0.22(0.03)&0.26(0.06)&95.60(2.15)&94.56(2.71)&0.85(0.14)&0.96(0.31)&98.04(0.33)&82.30(3.28)\\
&VoteNet$_2$+ (IS)&\textbf{0.21(0.04)}&\textbf{0.25(0.05)}&\textbf{96.29(1.82)}&\textbf{94.72(2.64)}&\textbf{0.82(0.22)}&\textbf{0.94(0.30)}&\textbf{98.12(0.28)}&\textbf{82.54(3.16)}\\
\hline
\hline
\multicolumn{1}{c}{\multirow{2}{*}{DIS}}&VoteNet$_0$+&0.24(0.03)&0.26(0.06)&94.64(2.21)&94.36(2.68)&0.90(0.16)&0.97(0.30)&97.89(0.33)&81.78(3.31)\\
&VoteNet$_1$+&0.22(0.03)&0.26(0.06)&95.43(2.19)&94.29(2.73)&0.85(0.15)&0.98(0.31)&98.01(0.33)&81.97(3.22)\\
\specialrule{.15em}{.05em}{.05em}
\end{tabular} 
\end{adjustbox}
\caption{Evaluation metrics for OAI segmentation performance. \textbf{ASD}: average surface distance. \textbf{SD}: average surface Dice score. \textbf{95MD}: 95 percentile of the maximum symmetric surface distance. \textbf{VD}: average volume Dice score. \textbf{$\uparrow$} (\textbf{$\downarrow$}) means the higher (lower) the better. \ding{\textbf{(I)} indicates registration refined using only $\mathcal{L}_{img}$ and $\mathcal{L}_{reg}$ while \textbf{(IS)} indicates also using $\mathcal{L}_{seg}$.}}
\label{tab:metrics}
\vspace{-0.5ex}
\end{table*}


\section{Experimental Results and Discussion}
\label{sec:experiment}
\subsection{Experimental Setting}
\label{sec:exp_setting}



\noindent
\textbf{Dataset.} For our experiments we use a 3D knee MRI dataset from the Osteoarthritis Initiative (OAI), which includes four labels (i.e., femur, tibia, femoral cartilage and tibial cartilage)~\cite{ambellan2019automated}. All images are affinely registered to an atlas built from training images. \ding{Images are resampled to size $160\times200\times200$ with isotropic spacing of 0.7 mm and intensity normalized to $[0, 1]$.} Our Train/Validate/Test split is 200/53/254 for both our registration networks and our U-Net~\cite{cciccek20163d} baseline segmentation network. To train VoteNet+, we randomly select 11 images (and their corresponding labels) from the training set as atlases and register these 11 atlases to the 200 training images to form the training dataset. 

\noindent
\textbf{Deep Networks.} For our pre-registration networks, we use the same architecture as in~\cite{xu2019deepatlas} to predict a displacement field, similar to~\cite{yang2017quicksilver,balakrishnan2018unsupervised}. 
\xu{(Quicksilver predicts the momentum. I think Deep atlas predcits a displacement field. So what do you mean by "predicting a disp field as in quicksilver?)\ding{see Marc's comment.}} 
\ding{All images are initially affinely registered to an atlas.} We train two models: one without label information as loss term (DI) and one with label information as loss term (DIS). We expect better accuracy from DIS as the network can benefit from stronger guidance via the given labels. We train using \texttt{ADAM} over 100 epochs with a multi-step learning rate. The initial learning rate is 0.0001 and reduced to 0.00001 after 50 epochs. We use soft multi-class Dice, normalized cross correlation, and the bending energy as loss terms. See~\cite{xu2019deepatlas} for details. We use a simple 3D U-Net~\cite{cciccek20163d} architecture for both U-Net segmentation and VoteNet+ probability prediction. 
For fair comparison, we train both models using \texttt{ADAM} with a cross entropy loss over 100 epochs with fixed learning rate 0.0001. VoteNet+ \mn{is trained based on the pre-registration results of the different registration models} to account for different registration accuracies. The models we consider are: VoteNet$_0$+ with \ding{only} Affine pre-registration, VoteNet$_1$+ with DI and VoteNet$_2$+ with DIS. It is expected that VoteNet$_2$+ is better than VoteNet$_1$+ and VoteNet$_0$+.
All models are trained on an NVIDIA RTX 8000 GPU.
\xu{(1. So which part is the VoteNet++? You need to define it somewhere. Is VoteNet+ plus DeepAtlas? Do you train them end-to-end? \ding{VoteNet++ is a symbol. we don't use deepatlas, we just use optimization to refine registration.} 2. What are VoteNet$_0$, VoteNet$_1$, VoteNet$_2$? I don't know the difference. For example VoteNe$_0$ with affine registration, is there no deformable registration involved at all? Also what is the registration model you use before the refinement network?)\ding{VoteNet$_0$ is trained with affine pre-registration; VoteNet$_1$ is trained with $V_1$ pre-registration; VoteNet$_2$ is trained with $V_2$ pre-registration.}}

\noindent
\textbf{Metrics.} We used 4 metrics to evaluate \mn{segmentation performance}: average surface distance, average surface Dice score (i.e., a surface element is considered overlapping if it is within a certain distance ($\leq$ 0.7 mm) to the other surface), 95\% maximum surface distance, and average volume Dice score.

\subsection{Performance Analysis}
Tab.~\ref{tab:metrics} shows our results. The plurality voting (PV) results show that the three registration models result in different segmentation performance; DIS is best, followed by DI, Affine shows the worst performance. All PV segmentation results are worse than the U-Net baseline. Using VoteNet+ for label fusion increases segmentation performance. In terms of volume Dice, we observe an average increase of 22\% for VoteNet$_0$+, 4\% for VoteNet$_1$+, and 0.5\% for VoteNet$_2$+ compared to PV. The other three metrics improve as well. Note that the performance of VoteNet$_2$+ is already on par with that of the U-Net baseline and even slightly better than it.

To test our hypothesis that \textit{a better registration model generally leads to the better VoteNet+ based MAS segmentation performance}, we use the DIS \ding{pre-}registration results to test VoteNet$_0$+ and VoteNet$_1$+. 
Note that DIS 
is expected to be better than Affine and DI \ding{pre-}registration. The bottom two entries of Tab.~\ref{tab:metrics} show the results. We observe that, after changing to a better \ding{pre-}registration model, DIS-VoteNet$_0$+ increases by 5.5\% volume Dice score on average over Affine-VoteNet$_0$+; DIS-VoteNet$_1$+ increases by 1.5\% volume Dice score on average over DI-VoteNet$_1$+. The other three metrics are also significantly improved. Note that DIS-VoteNet$_0$+ and DIS-VoteNet$_1$+ are comparable with the U-Net baseline, while the previous Affine-VoteNet$_0$+ and DI-VoteNet$_1$+ perform significantly worse than the U-Net baseline. This illustrates that registration refinement is promising in MAS. 

To refine imperfect registrations, we mainly use two loss terms as described in Sec.~\ref{sec:refine_regs}. $\mathcal{L}_{img}$ \mn{can be viewed as playing the same role as local patch search~\cite{coupe2011patch,wang2012multi,bai2013probabilistic,xie2019improving}, where $\mathcal{L}_{reg}$ controls spatial regularity}. 
Tab.~\ref{tab:metrics} shows that it improves segmentation performance significantly if the pre-registration model is not accurate (e.g. Affine-VoteNet$_0$+ (I) compared to Affine-VoteNet$_0$+); but it barely helps when the pre-registration model is accurate (e.g. DIS-VoteNet$_2$+ (I) compared to DIS-VoteNet$_2$). This \mn{may be} because the remaining anatomical difference tends to be large for inaccurate pre-registration algorithms and small for accurate pre-registration algorithms. Further, we introduce $\mathcal{L}_{seg}$ to help the registration refinement. Previous literature~\cite{xu2019deepatlas} has demonstrated that label information can improve registration accuracy. Inspired by this observation, we use the baseline U-Net to provide segmentation predictions for the warped atlas images and the target image. Since the predicted segmentations are from the same algorithm, they are expected to be consistent (i.e. simultaneously correct or wrong for the same type of anatomical structures). For inaccurate pre-registrations (i.e. Affine, DI), the performance improvement from VoteNet+ to VoteNet+ (IS) is very significant. For example, volume Dice increases by 3.9\% on average for Affine pre-registration and 1\% for DI pre-registration. 
For accurate pre-registration (i.e. DIS) we still observe moderate improvements. 
The volume Dice score increases by 0.08\% for bones and by 0.28\% for cartilage. \ding{Fig.~\ref{fig:result} demonstrates that our proposed refinement method makes registration more accurate and is better than local patch search.}

\ding{Furthermore, we find that pre-registration is vital for MAS. First, our pre-registration network can save computation time, because refinement based on good pre-registration results can converge more quickly. Second, refinement after poor pre-registration tends to result in worse accuracy than when starting with a good pre-registration. For example, in Tab.~\ref{tab:metrics}, Affine-VoteNet$_0$+ (IS) is worse than DIS-VoteNet$_0$+ and DI-VoteNet$_1$ (IS) is worse than DIS-VoteNet$_1$+. Hence, registration refinement is beneficial, but a good pre-registration is also important.}
\vspace{-0.5ex}
\section{Conclusion}
\label{sec:conclusion}

\ding{In this work, we studied registration refinement for multi-atlas segmentation. Specifically, we used predicted consistent segmentation information to improve atlas-to-target registration accuracy. We demonstrated that registration accuracy is vital for MAS and the predicted segmentation is beneficial for registration refinement.}


%
%
%
%
%

\pagebreak

\section{Compliance with Ethical Standards}
\label{sec:ethics}
This research study was conducted retrospectively using human subject data made available in open access form by the Osteoarthritis Initiative (OAI). Ethical approval was not required as confirmed by the license attached with the open access data.

\section{Acknowledgments}
\label{sec:acknowledgments}
Research reported in this work was supported by the National Institutes of
Health (NIH) and the National Science Foundation (NSF) under award numbers NSF EECS-1711776 and NIH 1R01AR072013. The content is solely the responsibility of the authors and does not necessarily represent the official views of the NIH or the NSF. The authors have no conflicts of interest.


\bibliographystyle{IEEEbib}
\bibliography{refs}

\end{document}